\documentclass[aps,prd,reprint,amsmath,amssymb,showpacs,nofootinbib]{revtex4-1}
\usepackage{epsfig,slashed,xcolor,bm,natbib,comment,enumitem,ulem,cancel}
\usepackage[cal=boondox]{mathalfa}
\def\bg#1{{\mbox{\boldmath $ #1 $}}}

\def\nd#1#2{{\frac{d #1}{d #2}}}

\newcommand{\comma}{, }
\newcommand{\be}{\begin{equation}}
\newcommand{\ee}{\end{equation}}
\newcommand{\bea}{\begin{eqnarray}}
\newcommand{\eea}{\end{eqnarray}}
\newcommand{\myDel}[1]{{\color{red}\ifmmode\cancel{#1}\else\sout{#1}\fi}}

\begin{document}
\title{New form of the Kerr--Newman solution} 
\author{M.P.~\surname{Hobson}}
\email{mph@mrao.cam.ac.uk} 
\affiliation{Astrophysics Group, Cavendish
  Laboratory, JJ Thomson Avenue, Cambridge CB3 0HE\comma UK}
%\date{Received 16 January 2020; accepted ???; published online ???}

\begin{abstract}
A new form of the Kerr--Newman solution is presented. The solution
involves a time coordinate which represents the local proper time for
a charged massive particle released from rest at spatial
infinity.  The chosen coordinates ensure that the solution is
well-behaved at horizons and enable an intuitive description of many
physical phenomena. If the charge of the particle $e = 0$, the
coordinates reduce to Doran coordinates for the Kerr solution with the
replacement $M \to M - Q^2/(2r)$, where $M$ and $Q$ are the mass and
charge of the black hole, respectively. Such coordinates are valid
only for $r \ge Q^2/(2M)$, however, which corresponds to the region that a
neutral particle released from rest at infinity can penetrate. By
contrast, for $e
\neq 0$ and of opposite sign to $Q$, the new coordinates have a
progressively extended range of validity as $|e|$ increases and tend
to advanced Eddington--Finkelstein (EF) null coordinates as $|e| \to
\infty$, hence becoming global in this limit. The Kerr solution
(i.e.\ with $Q=0$) may
also be written in terms of the new coordinates by setting $eQ =
-\alpha$, where $\alpha$ is a real parameter unrelated to charge;
in this case the coordinate system is global for all non-negative values of
$\alpha$ and the limits $\alpha = 0$ and $\alpha \to \infty$
correspond to Doran coordinates and advanced EF null coordinates,
respectively, without any need to transform between them.
\end{abstract}

%\pacs{04.50.Kd, 11.15.-q, 11.25.Hf}
%Modified theories of gravity, Gauge field theories, 
%Conformal field theory, algebraic structures
%11.30.Cp Lorentz and Poincaré invariance

\maketitle
%%%%%%%%%%%%%%%%%%%%%%%%%%%%%%%%%%%%%%%%%%%%%%%%%%%%%%%%%%%%%%%%%%%%%%%%%%

%\section{Introduction}
%\label{sec:intro}

The re-expression in different coordinate systems of exact solutions
in general relativity may at first seem a pointless exercise, since
the theory is covariant under such transformations by
construction. Nonetheless, a judicious choice of coordinates can
assist both in the mathematical calculation and physical intuition
associated with general-relativistic phenomena. Since one is often
interested in processes related to particle motion in the background
spacetime, it is thus natural to construct coordinate systems that are
based thereon.

In asymptotically flat black hole backgrounds, on which we will focus
our attention, consideration of ingoing or outgoing principal null
geodesics (which reduce to ingoing or outgoing radial photon
trajectories in the limit of a non-rotating black hole) lead to
advanced and retarded Eddington--Finkelstein (EF) coordinates,
respectively, which are well behaved at horizons and cover the entire
spacetime \cite{Eddington1924,Finkelstein1958}. Equally, one may
instead construct coordinate systems by considering the trajectories
of massive particles. The most natural of these are based on the
  motion of such particles released from rest at spatial infinity
  (which are often referred to as `raindrops') and, in particular,
  take the time coordinate $T$ (say) to represent the local proper
  time of the raindrop
  \cite{Painleve1921,Gullstrand1922,Novikov1963,Visser2005,Hamilton2008,Baines2021}.

For the Schwarzschild solution \cite{Schwarzschild1916}, demanding
that $\dot{T} = 1$, $\dot{\theta} = 0 = \dot{\phi}$ along a raindrop
trajectory (where a dot denotes the derivative with respect to the
raindrop proper time $\tau$) yields Painlev\'e--Gullstrand (PG)
coordinates \cite{Painleve1921,Gullstrand1922}, in which the metric
takes the form\footnote{We adopt the `mostly minus' metric signature
  $(+, -, -, -)$ and employ geometric units $c = G = 1$.}
\be
ds^2 = dT^2 - \left(dr + \sqrt{\frac{2M}{r}}\,dT \right)^2 -
r^2(d\theta^2 + \sin^2\theta\,d\phi^2),
\label{eqn:Schwarzschild-PG}
\ee
where $r$ lies in the range $0 < r < \infty$, and $\theta$ and $\phi$
take their usual meanings. Such coordinates are clearly global and
well-behaved at the horizon $r=2M$, unlike their standard
Schwarzschild counterparts, and so are useful in clarifying many
physical phenomena \cite{Visser2005, Nielsen2006, Abreu2010}, especially
black-hole thermodynamics \cite{Parikh2000,Vanzo2011}. In particular,
the 4-velocity of a raindrop in PG coordinates $x^\mu =
(T,r,\theta,\phi)$ has the simple form $\dot{x}^\mu =
(1,-\sqrt{2M/r},0,0)$.

Following an analogous procedure for the Reissner--Nordstr\"om (RN)
solution \cite{Reissner1916,Nordstrom1918}, one obtains a line-element
in PG coordinates again of the form (\ref{eqn:Schwarzschild-PG}), but
with $M \to M - Q^2/(2r)\equiv \mathcal{M}$, where $Q$ is the charge
of the black hole \cite{Nielsen2006,Abreu2010}. Although the PG
coordinates are again regular at horizons and useful for investigating
many physical phenomena, it is clear that the coordinates are not
global, since they are valid only for $r > Q^2/(2M)$
\cite{Lin2009}. This occurs because, in general, coordinates based on
the motion of some test particle are valid only in the region of the
spacetime that such a particle can penetrate, as is well known
\cite[e.g.][]{Hobson2005,Faraoni2020}; it is easily shown that $r=Q^2/2M$
marks the innermost radius that can be reached by a raindrop in the RN
solution.  As one might expect intuitively, for static,
spherically-symmetric spacetimes such as the RN solution, one can
address this shortcoming by instead constructing coordinates systems
based on the radial motion of massive particles with non-zero ingoing
coordinate speed at infinity $v_\infty$.  This yields the so-called
Martel--Poisson class of coordinates \cite{Martel2001}, which clearly
coincide with PG coordinates when $v_\infty = 0$, but have a
progressively larger range of validity as $v_\infty$ increases and
tend to advanced EF null coordinates as $v_\infty \to 1$, hence
becoming global in this limit.

Returning to PG coordinates, it is of particular note in
(\ref{eqn:Schwarzschild-PG}) that the metric coefficient $g_{00}$ (or
lapse function) is unity and the spacelike 3-surfaces $T =
\mbox{constant}$ are flat. Indeed, these
criteria comprise the original definition
\cite{Painleve1921,Gullstrand1922} of (strong) PG
coordinates\footnote{For weak PG coordinates, the lapse function is
  allowed to be non-trivial, while spacelike 3-surfaces are still
  required to be flat. One may also define conformal strong or weak PG
  coordinates, for which the spacetime line element is conformal to
  either the strong or weak versions of the PG form.}.  For the
Schwarzschild and RN solutions (and a number of other static,
spherically-symmetric spacetimes), the notions of a raindrop-based
time coordinate and a PG form for the metric coincide
\cite{Visser2005,Nielsen2006,Abreu2010,Faraoni2020}. Indeed, in such
cases, raindrops are often also called PG observers.

For rotating black holes, the situation is rather different.  Although
the Lense--Thirring solution \cite{Thirring1918,Mashoon1984} (the
slowly-rotating limit of the Kerr solution) can be expressed in
(strong) PG form
\cite{Baines2021a,Baines2021b,Baines2022a,Baines2022b}, the exact
Kerr(--Newman) metric cannot be put in strong, weak or even conformal
PG form \cite{Visser2022}. Nonetheless, one can
still construct useful coordinate systems for the Kerr(--Newman)
solution by again considering the motion of raindrops, although the link
between such coordinates and the PG form for the metric is broken in
this case.

The Kerr metric \cite{Kerr1963} in standard Boyer--Lindquist (BL)
coordinates \cite{Boyer1967} may be written as\footnote{This originally
  proposed form for the Kerr metric in BL coordinates \cite{Boyer1967}
  directly yields the form usually quoted \cite[e.g.][]{Hobson2005} on
  multiplying out and collecting coefficients of differentials.}
%
%\bea
%ds^2  &=& \left(1-\frac{2M r}{\rho^2}\right) \, dt^2 +
%\frac{4Mar\sin^2 \theta}{\rho^2} \, dt \, d\phi -
%\frac{\rho^2}{\Delta} \, dr^2 \nonumber \\
% &&-\rho^2 \, d\theta^2 
%\!\!-\!\! \left(r^2\!+\!a^2\!+\!\frac{2M ra^2\sin^2\theta}{\rho^2}\right)
%\sin^2\theta \, d\phi^2,
%\label{eqn:Kerr-BL}
%\eea
%
%
\bea
ds^2  &=&
\frac{\Delta}{\rho^2}(dt-a\sin^2\theta\,d\phi)^2 \nonumber \\
&& -\frac{\sin^2\theta}{\rho^2}[a\,dt - (r^2+a^2)\,d\phi]^2
-\frac{\rho^2}{\Delta}\,dr^2 -
\rho^2\,d\theta^2,\phantom{Aa}
\label{eqn:Kerr-BL}
\eea
where $\rho^2= r^2 + a^2\cos^2 \theta$ and $\Delta= r^2+a^2-2M r$. 
As will become clear from our further analysis
below, by considering the usual geodesic equations in BL
coordinates for a raindrop and introducing new coordinates $T$,
$\Theta$ and $\Phi$, for which one demands $\dot{T} = 1$,
$\dot{\Theta} = 0 = \dot{\Phi}$ (or, equivalently, $dT = d\tau$,
$d\Theta = 0 = d\Phi$) along a raindrop trajectory, one immediately
obtains $d\Theta = d\theta$ and 
\begin{subequations}
\label{eqn:Doran-diffs}
\bea
dT & = & dt + \frac{\sqrt{2Mr(r^2+a^2)}}{\Delta}\,dr,\\
d\Phi & = & d\phi + \frac{a}{\Delta}\sqrt{\frac{2Mr}{r^2+a^2}}\,dr.
\eea
\end{subequations}
In terms of these new coordinates (retaining the original
$\theta$-coordinate, which is clearly unchanged) the line element
(\ref{eqn:Kerr-BL}) may be written as
\bea
ds^2  &=& dT^2 - \left[\frac{\rho}{\sqrt{r^2+a^2}}\,dr +
  \frac{\sqrt{2Mr}}{\rho}(dT - a\sin^2\theta\,d\Phi) \right]^2
\nonumber \\
&& -\rho^2 \, d\theta^2 - (r^2 + a^2)\sin^2\theta\, d\Phi^2,
\label{eqn:Kerr-Doran}
\eea
which corresponds to Doran coordinates \cite{Doran2000}, although they
were not originally derived in the above manner.  It is clear that
this coordinate system, like PG coordinates for the Schwarzschild
solution, is both regular at horizons and global, and also useful for
describing many physical phenomena. In particular, the 4-velocity of a
raindrop in Doran coordinates $x^\mu = (T,r,\theta,\Phi)$ has the
simple form $\dot{x}^\mu =
(1,-\sqrt{2Mr(r^2+a^2)}/\rho^2,0,0)$. Various other coordinate systems
for the Kerr solution, based on raindrop trajectories or otherwise,
and their relationship to Doran coordinates are discussed in
\cite{Natario2009,Zaslavskii2018,Pavlov2019,Baines2021}. It is also
  worth noting that all these coordinate systems are related to
  advanced EF null coordinates by transformations of similar
  complexity to those in (\ref{eqn:Doran-diffs}).

The Kerr-Newman (KN) solution \cite{Newman1965,Adamo2014} is the most
general asymptotically-flat, stationary solution of the
Einstein--Maxwell equations in general relativity.  By analogy with
the transition from the Schwarzschild to RN solution, the standard KN
metric in BL coordinates may be obtained from (\ref{eqn:Kerr-BL}) by
again making the substitution $M \to M - Q^2/(2r) \equiv \mathcal{M}$
throughout. Similarly, the transition to the KN solution in Doran
coordinates is also immediately obtained by setting $M \to
\mathcal{M}$ in (\ref{eqn:Kerr-Doran})
\cite{Hamilton2008,Jiang2006}. For the KN solution, however, it is
clear that one encounters an analogous difficulty to that discussed
earlier for the RN solution, namely that the Doran coordinates are not
global since they are valid only for $r > Q^2/(2M)$ \cite{Lin2009};
this is again easily shown to correspond to the innermost radius to
which a raindrop trajectory can penetrate.

By analogy with the RN solution, one might hope to address this
shortcoming by constructing coordinate systems based on the motion of
massive particles with non-zero ingoing coordinate speed at infinity
$v_\infty$, which should penetrate to smaller values of $r$, thereby
extending the notion of the Martel--Poisson class of coordinate
systems to rotating black hole backgrounds. While this approach is
valid, in principle, we find that it leads to a rather cumbersome
and unintuitive set of coordinate systems that moreover do not tend to
advanced EF null coordinates in the limit $v_\infty \to 1$. We
therefore instead consider the alternative approach of constructing a
coordinate system based on the motion of a {\it charged} raindrop. The
basic intuition in so doing is that a raindrop with charge of opposite
sign to that of the black hole experiences an additional inwards
electromagnetic force that should allow it to penetrate to smaller
values of $r$ than a neutral raindrop.

We begin our analysis by considering the orbit equations in BL
coordinates for a particle of unit mass and charge $e$ in the KN
spacetime, which read \cite{Carter1968}
\begin{subequations}
\label{eqn:KNgeodesics}
\bea 
\rho^2 \dot{t} & = & -a(ak\sin^2\theta-h)
+\frac{r^2+a^2}{\Delta}P(r), \\ 
\rho^2 \dot{r} & = & \pm
\sqrt{P(r)^2 - \Delta[r^2 + (h-ak)^2 + C]},\phantom{AA}\\ 
\rho^2 \dot{\theta} & = & \pm \sqrt{C - \cos^2\theta\left[a^2(1-k^2) +
    \frac{h^2}{\sin^2\theta}\right]}\\ 
\rho^2 \dot{\phi} & = &
-\left(ak-\frac{h}{\sin^2\theta}\right) + \frac{a}{\Delta}P(r), 
\eea
\end{subequations}
where $P(r) \equiv k(r^2+a^2)-ah-eQr$ and the orbit is
characterised by the particle specific energy $k$, specific angular
momentum $h$ and specific Carter's constant $C$, all of which
are conserved quantities.

For a raindrop, the conserved quantities along the orbit have the
values $k=1$, $h=0=C$. In this special case, the orbit equations
(\ref{eqn:KNgeodesics}) are greatly simplified. In particular, one
sees that this case is unique in yielding $\dot{\theta}=0$.  Moreover,
one obtains
\begin{subequations}
\label{eqn:chargetrajectory}
\bea
\nd{t}{r} & = & \frac{\dot{t}}{\dot{r}} 
= -\frac{(\rho^2-eQr)(r^2+a^2) 
+ 2\mathcal{M}ra^2\sin^2\theta}{\Delta\sqrt{R(r)}},\phantom{AA}\\
\nd{\phi}{r} & = & \frac{\dot{\phi}}{\dot{r}} =
-\frac{(2\mathcal{M}-eQ)ra}{\Delta\sqrt{R(r)}},
\eea
\end{subequations}
where we have assumed that the raindrop is
ingoing and have defined the cubic $R(r) \equiv
2(\mathcal{M}-eQ)r(r^2+a^2)+e^2Q^2r^2$. If one now introduces new coordinates $T$, $\Theta$ and
$\Phi$, for which one demands $\dot{T} = 1$, $\dot{\Theta} = 0 =
\dot{\Phi}$ (or, equivalently, $dT = d\tau$, $d\Theta = 0 = d\Phi$)
along a raindrop trajectory, one immediately obtains $d\Theta =
d\theta$ and
\begin{subequations}
\label{eqn:newcoords-diffs}
\bea
dT & = & dt + \frac{r^2+a^2}{\Delta}F(r)\,dr\\
d\Phi & = & d\phi + \frac{a}{\Delta}F(r)\,dr.
\eea
\end{subequations}
where we have defined $F(r) = (2\mathcal{M}-eQ)r/\sqrt{R(r)}$.  

Before rewriting the KN metric in terms of these coordinates, it is
useful to consider some limits of the above transformations. As
expected, if $e=0$ then $R(r) = 2\mathcal{M}r(r^2+a^2)$ and $F(r) =
\sqrt{2\mathcal{M}r/(r^2+a^2)}$, such that the transformations
(\ref{eqn:newcoords-diffs}) reduce directly to those in
(\ref{eqn:Doran-diffs}) with the replacement $M \to \mathcal{M}$,
which define Doran coordinates for the Kerr--Newman solution that are
valid for $r > Q^2/(2M)$; clearly these coordinates further reduce to
the original globally-valid Doran coordinates for the Kerr solution if
$Q=0$. If $e \neq 0$ and is of opposite sign to $Q$, the coordinates
defined by the transformations (\ref{eqn:newcoords-diffs}) enjoy a
progressively extended range of validity as $|e|$ increases. The
innermost value of $r$ for which the coordinates are valid is clearly
given by the single real root of the cubic $R(r)$; the corresponding
expression for this root is rather cumbersome and unenlightening, but
always lies in the range $0 < r < Q^2/(2M-eQ)$. In the limit $|e| \to
\infty$ (again with $eQ < 0$), $F(r) \to 1$ and so the transformations
(\ref{eqn:newcoords-diffs}) tend to those defining advanced EF null
coordinates (where $dT$ is usually denoted by $du$)
\cite{Carter1968}. Indeed, from (\ref{eqn:chargetrajectory}) in the
same limit, one finds that the particle trajectory coincides will the
principal null geodesics, along which $dT = 0$, as expected. In this
limit, the coordinate system thus becomes global.

In terms of the new coordinates (\ref{eqn:newcoords-diffs}), the KN
solution, which is given in BL coordinates by (\ref{eqn:Kerr-BL}) with
$M \to \mathcal{M}$, becomes
\vspace*{-2mm}
\bea
ds^2  &=& \frac{\Delta}{\rho^2}\left[\frac{\rho^2}{\Delta}F(r)\,dr-
(dT-a\sin^2\theta\,d\Phi)\right]^2 \nonumber \\
&&-\frac{\sin^2\theta}{\rho^2}[a\,dT \!-\! (r^2+a^2)\,d\Phi]^2
\!-\!\frac{\rho^2}{\Delta}\,dr^2 \!-\!
\rho^2\,d\theta^2,\phantom{Aa}
\label{eqn:Kerr-new}
\\\nonumber
\eea
which in fact holds for arbitrary $F(r)$ in
(\ref{eqn:newcoords-diffs}) (other forms for $F(r)$ are
considered in \cite{Rosquist2009}). An equivalent form of the KN
solution in the new coordinates is given by
\begin{widetext}
\bea ds^2 & = & dT^2 - \frac{2(\mathcal{M}-eQ)r\rho^2}{R(r)}
\left[dr+\frac{\sqrt{R(r)}}{\rho^2}\,(dT-a\sin^2\theta\,d\Phi)\right]
\left[dr+\frac{\mathcal{M}}{\mathcal{M}-eQ}\frac{\sqrt{R(r)}}{\rho^2}
\,(dT-a\sin^2\theta\,d\Phi)\right] \nonumber \\
&& - \rho^2\,d\theta^2 - (r^2+a^2)\sin^2\theta\,d\Phi^2,
\label{eqn:Kerr-final}
\eea
\end{widetext}
which may be more useful in some physical applications.  In
particular, one sees immediately that for a charged raindrop, for
which $\rho^2\,dr = -\sqrt{R(r)}\,dT$ and $d\theta = 0 = d\Phi$, the
line element reduces to $ds^2 = dT^2$, as required, so the raindrop
4-velocity is $\dot{x}^\mu = (1,-\sqrt{R(r)}/\rho^2,0,0)$.  It is
further worth noting that $ds^2 = dT^2$ {\it also} holds for particles
for which $\rho^2\,dr = -\mathcal{M}\sqrt{R(r)}\,dT/(\mathcal{M}-eQ)$
and $d\theta = 0 = d\Phi$.

For completeness, we note that the electromagnetic vector potential of
the KN solution in
the new coordinate system is trivially obtained from its standard form
in BL coordinates, most succinctly expressed as the 1-form $A =
A_\mu\,dx^\mu = (Qr/\rho^2)(dt-a\sin^2\theta\,d\phi)$, by using the
transformations (\ref{eqn:newcoords-diffs}) to obtain
\be
A = \frac{Qr}{\rho^2}\left[dT - \frac{\rho^2}{\Delta}F(r)\,dr 
- a\sin^2\theta\,d\Phi\right].
\ee

For $e=0$, the line element (\ref{eqn:Kerr-final}) is easily shown to
reduce to the Doran form (\ref{eqn:Kerr-Doran}) with $M \to \mathcal{M}$,
as expected.  As also anticipated, in the limit $|e| \to \infty$ (with
$eQ < 0$), for which $F(r) \to 1$, the line element
(\ref{eqn:Kerr-new}) is quickly found to reduce to the KN solution in
advanced EF null coordinates (where $dT$ is usually denoted by $du$),
namely
\bea
ds^2 & = & \left(1-\frac{2\mathcal{M}r}{\rho^2}\right)\,dT^2 -2\,dT\,dr +
\frac{4\mathcal{M}ra\sin^2\theta}{\rho^2}\,dT\,d\Phi \nonumber \\
&& +2a\sin^2\theta\,dr\,d\Phi  -\rho^2\,d\theta^2 \nonumber \\ 
&& -\left(
r^2+a^2 +\frac{2\mathcal{M}ra^2\sin^2\theta}{\rho^2}
\right)\sin^2\theta\,d\Phi^2.
\eea

It is worth pointing out that the Kerr solution (for which $Q=0$) may
also be written in terms of the new coordinate system by setting
$\mathcal{M} = M$ and $eQ = -\alpha$ throughout (\ref{eqn:Kerr-final}),
where $\alpha$ should now be considered simply as a real parameter
that is unrelated to charge. In this case, the coordinate system is
regular at horizons and global for any non-negative value of $\alpha$, and the
limits $\alpha = 0$ and $\alpha \to \infty$ correspond to Doran
coordinates and advanced EF null coordinates, respectively, without
any need to transform between them.

Finally, one should note that the trajectories of charged or neutral
massive test particles, as usually considered, are somewhat
hypothetical compared to the motion of physical massive particles
falling into a black hole. In particular, for a charged particle one
should take into account the back reaction on the particle resulting
from the emission of electromagnetic radiation along its trajectory
\cite[e.g.][]{Folacci2020,Komarov2022}. Even for a neutral particle,
one should in principle consider the back reaction due to the emission
of gravitational radiation, although this will be a much smaller
effect. It is worth noting, however, that tidal forces on an infalling
physical neutral massive particle will in any case eventually tear it apart
into its charged fundamental constituents (unless the original neutral
massive particle is a Higgs boson, which is the only such fundamental particle
in the Standard Model). Thus, the 4-velocity of a physical
massive particle in a black-hole background will not, in general, take a
simple form in any coordinate system based on the idealised motion of
test particles.

%%%%%%%%%%%%%%%%%%%%%%%%%%%%%%%%%%%%%%%%%%%%%%%%%%%%%%%%%%%%%%%%%%%%%%%%%%%%%
%\medskip
\vspace{-0.2cm}
\begin{acknowledgments}
\vspace*{-0.25cm} 
The author thanks Anthony Lasenby for
helpful discussions and the anonymous referee for some useful suggestions.
\end{acknowledgments}
%%%%%%%%%%%%%%%%%%%%%%%%%%%%%%%%%%%%%%%%%%%%%%%%%%%%%%%%%%%%%%%%%%%%%%%%%%


\begin{thebibliography}{99}

\bibitem{Eddington1924}
A.S. Eddington, Nature (London) {\bf 113}, 192 (1924).

\bibitem{Finkelstein1958}
D. Finkelstein, Phys. Rev. D {\bf 110}, 965 (1958).

\bibitem{Painleve1921}
P. Painlev\'e, C. R. Acad. Sci. (Paris) {\bf 173}, 677 (1921).

\bibitem{Gullstrand1922}
A. Gullstrand, Arkiv. Mat. Astron. Fys. {\bf 16}, 1 (1922).

\bibitem{Novikov1963}
I.D. Novikov, Doctoral dissertation, Shternberg Astronomical
Institute, Moscow, 1963.

\bibitem{Visser2005}
M. Visser, Int. J. Mod. Phys. D {\bf 14}, 2051 (2005).

\bibitem{Hamilton2008}
A. Hamilton and J. Lisle, Am. J. Phys. {\bf 76}, 519 (2008).

\bibitem{Baines2021}
J. Baines, T. Berry, A. Simpson and M. Visser,
Classical Quantum Gravity {\bf 38}, 055001 (2021).

\bibitem{Schwarzschild1916} K. Schwarzschild,
  Sitzungsber. Preuss. Akad. Wiss. Berlin (Math. Phys.), 189 (1916).

\bibitem{Nielsen2006}
A.B. Nielsen and M. Visser, Classical Quantum Gravity {\bf 23}, 4637 (2006).

\bibitem{Abreu2010}
G. Abreu and M. Visser, Phys. Rev. D {\bf 82}, 044027 (2010).

\bibitem{Parikh2000}
M.K. Parikh and F. Wilczek, Phys. Rev. Lett. {\bf 85}, 5042 (2000).

\bibitem{Vanzo2011} 
L. Vanzo, G. Acquaviva and R. Di Criscienzo, Classical Quantum Gravity {\bf
  28}, 183001 (2011).

\bibitem{Reissner1916} 
H. Reissner, Ann. Phys. (Berlin) {\bf 50}, 106 (1916).

\bibitem{Nordstrom1918} G. Nordstr\"om,
  Verhandl. Koninkl. Ned. Akad. Wetenschap., Afdel. Natuurk.,
  Amsterdam {\bf 26}, 1201 (1918).

\bibitem{Lin2009}
C.-Y. Lin and C. Soo, Phys. Lett. B {\bf 671}, 493 (2009).

\bibitem{Hobson2005}
M.P. Hobson, G.P. Efstathiou, and A.N. Lasenby, {\it General
Relativity: An Introduction for Physicists} (Cambridge
University Press, Cambridge, England, 2005).

\bibitem{Faraoni2020}
V. Faraoni and G. Vachon, Eur. Phys. J. C {\bf 80}, 771 (2020). 

\bibitem{Martel2001}
K. Martel and E. Poisson, Am. J. Phys. {\bf 69}, 476 (2001).

\bibitem{Thirring1918}
H. Thirring and J. Lense, Phys. Z. Leipzig Jg. {\bf
  19}, 156 (1918).

\bibitem{Mashoon1984}
B. Mashoon, F.W. Hehl and D.S. Theiss, Gen. Relativ. Gravit. {\bf 16}, 711 (1984).

\bibitem{Baines2021a}
J. Baines, T. Berry, A. Simpson, and M. Visser,
Universe {\bg 7}, 105 (2021).

\bibitem{Baines2021b}
J. Baines, T. Berry, A. Simpson, and M. Visser,
Universe {\bg 7}, 473 (2021).

\bibitem{Baines2022a}
J. Baines, T. Berry, A. Simpson, and M. Visser,
Universe {\bg 8}, 115 (2022).

\bibitem{Baines2022b}
J. Baines, T. Berry, A. Simpson, and M. Visser,
Gen. Relativ. Gravit. {\bf 54}, 79 (2022).

\bibitem{Visser2022}
M. Visser and S. Liberati, Gen. Relativ. Gravit. {\bf 54}, 145 (2022).

\bibitem{Kerr1963}
R.P. Kerr, Phys. Rev. Lett. {\bf 11}, 237 (1963).

\bibitem{Boyer1967}
R.H. Boyer and R.W. Lindquist, J. Math. Phys. (N.Y.) {\bf 8}, 265 (1967).

\bibitem{Doran2000}
C. Doran, Phys. Rev. D {\bf 61}, 067503 (2000).

\bibitem{Natario2009}
J. Nat\'ario, Gen. Relativ. Gravit. {\bf 41}, 2579 (2009).

\bibitem{Zaslavskii2018}
O.B. Zaslavskii, Gen. Relativ. Gravit. {\bf 50},  123 (2018).

\bibitem{Pavlov2019}
Y.V. Pavlov and O.B. Zaslavskii, Gen. Relativ. Gravit. {\bf 51}, 60 (2019).

\bibitem{Newman1965}
E.T. Newman, R. Couch, K. Chinnapared, A. Exton, A. Prakash and
R. Torrence, J. Math. Phys. (N.Y.) {\bf 6}, 918 (1965).

\bibitem{Adamo2014}
T. Adamo and E.T. Newman, Scholarpedia {\bf 9}, 31791 (2014).

\bibitem{Jiang2006}
Q.Q. Jiang, S.-Q. Wu and X. Cai, Phys. Rev. D {\bf 73}, 064003 (2006).

\bibitem{Carter1968}
B. Carter, Phys. Rev. {\bf 174}, 1559 (1968).

\bibitem{Rosquist2009}
K. Rosquist, Gen. Relativ. Gravit. {\bf 41}, 2619 (2009).

\bibitem{Folacci2020}
A. Folacci and M. Ould El Hadj, Phys. Rev. D {\bf 102}, 024026 (2020).

\bibitem{Komarov2022} S.O. Komarov, A.K. Gorbatsievich, A.S. Garkun and
  G.V. Vereshchagin, arXiv:2211.04544.


\end{thebibliography}
\end{document}